\documentclass[11pt]{article}
\usepackage{amsfonts}
\usepackage{amsmath}
\usepackage{amssymb}
\usepackage{amsbsy}
\usepackage[T1]{fontenc}
\usepackage[british]{babel}
\usepackage{latexsym}
\usepackage{mathtools}
\usepackage{setspace}
\usepackage{caption}
\usepackage{color}   
\usepackage{epsfig}
\usepackage{graphics}
\usepackage{graphicx}
\usepackage{xmpmulti}
\usepackage[varg]{txfonts}
\usepackage{aas_macros} 
\usepackage{natbib} 
\def\i{\item}

\newcommand{\bed}{\begin{displaymath}}
\newcommand{\eed}{\end{displaymath}}
\newcommand{\bei}{\begin{itemize}}
\newcommand{\eei}{\end{itemize}}
\newcommand{\bef}{\begin{figure}}
\newcommand{\eef}{\end{figure}}
\newcommand{\ben}{\begin{enumerate}}
\newcommand{\een}{\end{enumerate}}
\newcommand{\beq}{\begin{equation}}
\newcommand{\eeq}{\end{equation}}
\newcommand{\ber}{\begin{eqnarray}}
\newcommand{\eer}{\end{eqnarray}}

\begin{document}

\title{Under-Graduate Research in Physics : An Indian Perspective}
\author{Sushan Konar, NCRA-TIFR, Pune, India \\
{\em email : sushan@ncra.tifr.res.in}
}
\maketitle

\begin{abstract}
  It is now  widely believed that research should be  an essential and
  integral part of  under-graduate studies. In recent  years there has
  been  a conscious  effort  to bring  research  opportunities to  the
  physics under-graduates  in India.  We argue that  the need  for the
  hour  is  a methodical  evaluation  of  the existing  under-graduate
  research programs for  their effectiveness  in preparing  the students
  for a career in physics.
\end{abstract}

\section*{Introduction}
Following   the  Boyer   Commission  Report   (1998)  and   subsequent
recommendations by a  number of academic bodies,  the higher education
policy in the US underwent a paradigm shift by recognising research as
an essential and  integral part of the  Under-Graduate (UG) education.
A  number of  investigations have  already demonstrated  the range  of
personal, professional  and intellectual benefits that  STEM (Science,
Technology,   Engineering   \&   Mathematics)   students   gain   from
participating in Under-Graduate Research (UGR) \cite{crowe08}.

Of  late, we  have observed  a similar  paradigm shift  in the  Indian
context too. An UG student had little access to research just a couple
of decades ago.   However, as the benefits of UGR  become apparent, in
terms of  students' learning retention, graduation  rates and entrance
into  graduate  programs~\cite{russell07,wilke07},  there has  been  a
change in the higher-education  policy supplemented by generous grants
from funding agencies.  This has been hailed enthusiastically, both by
practising scientists  as well as the  student community, particularly
because   UGR  is   increasingly  being   considered  as   a  critical
qualification for graduate admissions abroad as well as in India.

Recent studies also indicate that the growth of physics as a field and
the retention of  students desirous of being physicists  is the lowest
among  all   STEM  subjects~\cite{irvin15}.    It  appears   that  the
development  of  a  professional  identity is  important  for  student
retention and UGR can  play a major role in this  direction. An UGR is
where a student realises for the first time that the classroom setting
provides answers  while the research  experience focuses on  asking as
well as  answering questions.   This ultimately  helps UG  students to
`become  scientists'  through  the  growth and  development  of  their
professional identities~\cite{hunte07}.   Therefore scholarly studies,
defining and measuring the metric against which the UGR experience can
be  evaluated,  is of  great  importance  in  the context  of  physics
research in  India.  This essay argues  for a need of  such studies by
considering the  situation of the  physics UGR opportunities  that are
currently available in the country.

\section*{UGR (Physics) in India}

There  are  basically  four  types of  institutions  (leaving  defence
laboratories aside) where physics research is conducted in India. They
could be classified as follows -
\ben
   \i  the  traditional  universities  with  a  number  of  affiliated
   colleges (Bombay, Calcutta, Delhi, Madras, Pune etc.);
   \i   the  stand-alone   autonomous  universities   (Cotton  College
   (Guwahati),   Jadavpur   (Kolkata),  JNU   (Delhi),   Visva-Bharati
   (Shantiniketan) etc.);
   \i  the specialised  STEM-only universities  (IISc (Bangalore),  the
   IITs, the IISERs, BITS etc.); and
   \i the elite research  institutes (HRI (Allahabad), IMSc (Chennai),
   PRL(Ahmedabad), RRI (Bangalore), TIFR (Mumbai), the IUCs etc.).
\een
While  the  emphasis is  on  UG  teaching  (UGT)  in the  first  three
categories,   the  primary   activity   is  pure   research  for   the
fourth. However, with  the introduction of Integrated  PhD programs, a
number of  research institutes (like  HRI, IMSc, TIFR etc.)   are also
engaging in  UGT now.  Even though  the prevalence of UGR  is a recent
phenomenon,  the  understanding  of   the  importance  of  a  research
component in the  UG curriculum and a conscious  effort to incorporate
such  an element  in  the Indian  system has  existed  for quite  some
time.  Initially,   the  efforts  were   restricted  only  to   a  few
universities (mostly IITs)  but now students from  almost all academic
institutions  can access  research opportunities  in one  form or  the
other.  Interestingly,  these  UGR   opportunities  belong  to  a  few
distinctive  types  (see  Fig.\ref{ugr_type})  which  can  be  broadly
classified under the  following two categories -  {\em curricular} and
{\em extra-curricular}.

\bef
  \begin{center}
    \includegraphics[width=12.0cm]{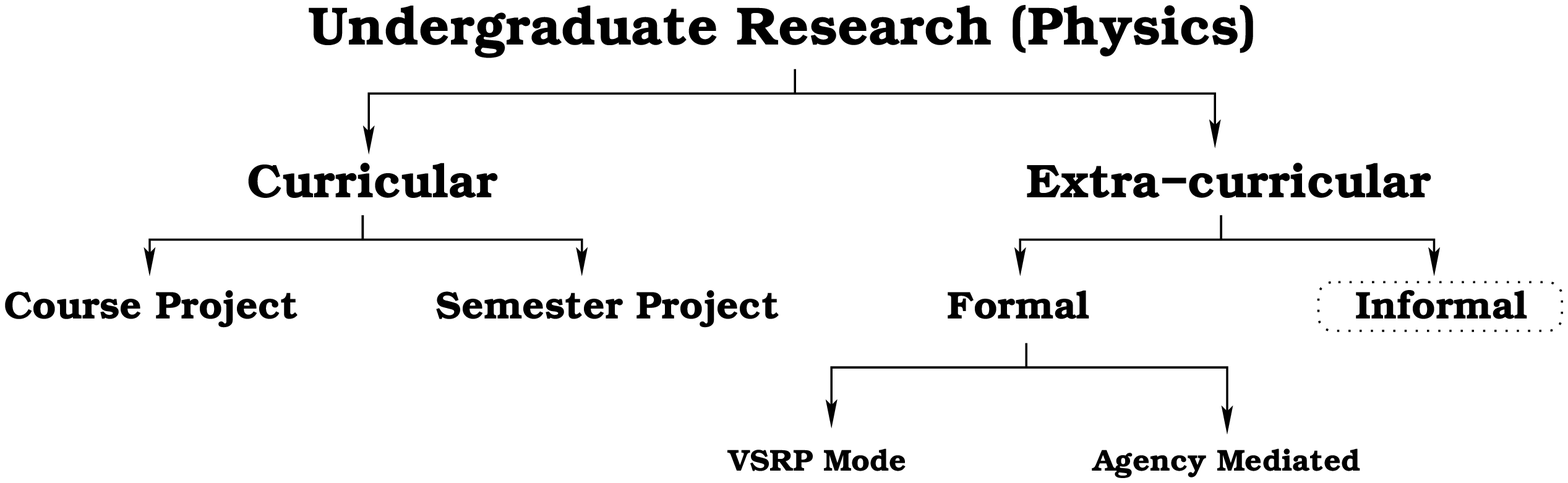}
  \end{center}
\caption[]{Types of  physics UGR that  are currently being  offered in
  India.}
\label{ugr_type}
\eef

\subsection*{Curricular UGR}
A curricular UGR project is a  piece of research undertaken to fulfil
the  requirements  of  the  university syllabus  where  a  student  is
enrolled  for  her/his UG  (bachelors/masters)  degree.   One type  of
curricular research that is becoming  increasingly popular is the very
short-term {\em course project} - forming a part of a standard course.
Usually  such  a  project  is  undertaken  in  lieu  of  the  end-term
examination and may end up being nothing more than a longish take-home
examination.  However, depending on the situation such a project could
also involve reading a research  article, understanding it and perhaps
taking  some  of the  calculations  a  little forward.   Though  these
typically do not  produce any serious research output,  they offer the
students a  quick look at the  methodology of research.  On  the other
end  of spectrum  of  curricular research  is  the {\em  semester-long
  project},   where   the  duration   could   even   span  an   entire
year. Depending  on the  university, a student  is either  required to
work with a professor in the home institution (sometimes continuing to
do other course-work alongside); or is given the option to work with a
scientist  from another  academic institute.   Because of  the longish
duration,  a student  usually gets  sufficient  time to  pick up  some
background     and/or     acquire     certain     necessary     skills
(computational/experimental/instrumental)  before starting  the actual
work.

The single-most important feature of these curricular projects is that
a  degree  of  sincerity  and seriousness  is  automatically  ensured,
because  the  research  performance  gets  reflected  in  a  student's
graduating  grade  which  plays  a decisive  role  in  the  subsequent
placement of the  student. Consequently, good amount  of research work
come  out  of  these.   Oftentimes,  an important  part  of  a  larger
investigation (theoretical/experimental/observational) is successfully
set-up by  an UG  student irrespective  of whether or  not it  ends up
being written  up as a  full-fledged publication.  At GMRT  (the Giant
Meterwave  Radio  Telescope, maintained  by  NCRA-TIFR,  Pune) a  good
amount of engineering  and software support has, over  the years, been
generated    through     such    research    projects     (see,    for
example~\cite{de16}).  Overall,  the impact  on and output  from these
projects on UG learning appear to be quite positive.

Most  of  the  `stand-alone'  and `STEM-only'  universities  have  now
incorporated  such research  projects (both  short \&  long-term) into
their UG programs.  However, traditional universities are yet to adopt
these measures and the research  component is still mostly absent from
their curricula.

\subsection*{Extra-curricular UGR}
In  contrast,  an extra-curricular  research  is  something a  student
embarks  upon  of  his  own  volition.  The  motivation  for  such  an
undertaking has a wide range -  from a genuine desire for research, to
securing  an early  admission to  a prestigious  graduate program,  or
perhaps to simply affect an  improvement to the bio-data.  These could
be {\em  formal}, structured by  particular academic institutes  or by
umbrella organisations  (like the Indian Science  Academies).  By far,
the best-known and likely the oldest running program in this category,
is the `Visiting  Students Research Program' ({\em VSRP})  run by TIFR
(similar  programs  are  now  being offered  by  many  other  research
institutes). Applicants from across the country are filtered through a
stiff selection  criterion. Upon being  selected a student works  on a
research  project,  typically  lasting  for 8-10  weeks,  under  close
supervision of a faculty member.  The highlight of a such a program is
that at  the end of  their stay the  students are evaluated  for their
suitability  to  join the  research  program  of that  institute  upon
completing their respective UG degrees.

A new addition  to the formal, extra-curricular  research is something
we  term  as  an  {\em  agency  mediated}  one.   The  NIUS  (National
Initiative on Undergraduate Science) program  run by HBCSE-TIFR or the
SRFP  (Summer Research  Fellowship Program)  jointly sponsored  by the
three science  academies (Indian Academy of  Sciences, Indian National
Science Academy \&  The National Academy of Sciences)  are examples of
this  type. Once  again students  are selected  from a  huge applicant
pool.   Afterwards they  are assigned  to faculty  mentors working  in
different  academic  institutes  across the  country.   The  mediating
agency basically pairs up the  aspiring UG student with an appropriate
mentor  and fund  the  entire activity.   Though  these allow  immense
structural  freedom  (the student  and  the  supervisor can  choose  a
convenient time-frame or the number of visits the student makes to the
host institute and so on)  conducive to serious research, the programs
may  not always  achieve  it.  One of  the  reasons  could be  certain
shortcomings  of the  selection  procedure. With  a completely  uneven
grading  system  of our  universities,  the  probability of  selecting
students with  inadequate background  training or motivation  is quite
high when the selection depends  mainly on their university grades. In
contrast, the probability  of making similar mistakes  is much smaller
in similar programs  run by individual institutes where  the number of
selected applicants is usually quite  modest.  Moreover, a fraction of
students  also consider  these  programs simply  as opportunities  for
improving  the bio-data.   In such  cases, non-performance  is not  an
issue unless  and until the  student requires a reference  letter from
the faculty-mentor.

In  fact, this  last point  brings us  to the  emergent trend  of {\em
  informal} research projects. All of the UGR that have been discussed
so far are `structured'  programs, conducted by academic institutions.
However, the  need for an impressive  bio-data and a set  of reference
letter  writers  are now  compelling  students  to informally  contact
potential mentors  looking for research  opportunities. To be  sure, a
fraction  of   these  requests   arise  out  of   genuine  motivation.
Sometimes, a student who have  earlier worked with a particular mentor
(through another form of UGR)  returns to complete the work, resulting
in  a   publication  (for  example,   see~\cite{konar04c}).   Research
institutes  usually provide  infrastructural support  and funding  for
these if a faculty member agrees  to mentor a student.  These projects
can  range  from  a few  weeks  to  even  a  year depending  upon  the
situation.  Excellent  pieces of research  work have come out  of such
projects.   However,  because  many  of  these  project  requests  are
initiated by students who have been unable to secure a place in one of
the  above-mentioned formal  programs,  it appears  that  a number  of
disturbing issues have become associated  with this particular form of
UGR.

Unfortunately,  the requests  for informal  projects has  reached such
proportions that scientists  working both within India  and abroad are
getting completely overwhelmed  by the huge volume  of unsolicited and
mostly irrelevant emails.  It is learnt that in  many universities the
student communities have developed computer scripts by which they have
automated  the process  of  selecting potential  mentors (by  trawling
through  internet  repositories  of scholarly  articles  etc.).  These
requests should really be considered as  `spam's which hope to find at
least  one gullible  person  somewhere  and are  driven  by the  sheer
desperation of students.   To be honest, this craze  is also nurtured,
in part,  by individuals or  groups that require a  large `non-expert'
workforce to help conduct big  experiments, or analyse huge volumes of
data, or run monstrous simulations.

\section*{Concerns}
Easy access to UGR opportunities for  a large UG community is a recent
phenomenon in India. Evidently, there are quite a few issues that need
to be looked  into.  Clearly, the older and  structured programs (like
the VSRP) do better than the more recent programs.  However, it is not
yet clear  if every type of  UGR is as effective  as naively expected.
For example, it is not uncommon to see students going through a string
of short-term UGRs  on totally unrelated topics and  gaining neither a
feel for the methodology nor any real training in a particular area of
research.  The unduly excessive emphasis  placed on the UGR experience
for admissions  into graduate programs  is likely the reason  for this
obsession, oftentimes even to the exclusion of any physics learning.

The selection  of under-prepared and/or under-motivated  students to a
UGR program  is another source of  serious problem. As this  is mainly
associated  with agency-mediated  programs  where  the applicant  pool
could be  enormous a  better methodology  need to  be adopted  for the
selection  process.   Introduction of  a  `tiered  structure' to  such
programs could also be thought of,  where the present structure can be
followed  for  students  properly   equipped  for  research;  and  the
under-prepared students  could be  inducted into `summer  school' like
programs  where they  reinforce  their UG  learning  itself. The  REAP
(Research Education Advancement Programme) run by the Jawaharlal Nehru
Planetarium (Bangalore) is an example of such a program.

To conclude,  it has  to be said  that even though  the inputs  to the
present essay are  mostly anecdotal there is a clear  indication for a
need for quantitative studies on the effect of UGR on physics research
in India. Whether the resource spent (in terms of person-hours, direct
funding  and   infrastructural  support)   is  actually   helping  the
indigenous research community or is  simply producing students who are
better equipped  to ensure admission  in a graduate program  abroad is
something we seriously need to look into.

The author is indebted to D. Narasimha for setting a gold standard for
an UGR supervisor when she worked with  him for a VSRP project at TIFR
some three  decades ago; also  to Niruj Mohan Ramanujan  \& Tirthankar
Roy  Choudhury for  providing her  with certain  valuable inputs.  The
author  is  supported  by  a grant  (SR/WOS-A/PM-1038/2014)  from  the
Department of Science \& Technology, Government of India.

\bibliographystyle{pramana}
\bibliography{adsrefs}

\end{document}